\documentclass[prl,aps,amssymb,twocolumn,showpacs,floatfix]{revtex4}
\usepackage[dvips]{graphicx}
\def\be{\begin{equation}}
\def\ee{\end{equation}}
\def\bea{\begin{eqnarray}}
\def\eea{\end{eqnarray}}

\begin{document}

\title{Exact relations between multifractal exponents at the Anderson
 transition}
\author{A.D.~Mirlin$^{1,2,*}$}
\author{Y.V.~Fyodorov$^3$}
\author{A.~Mildenberger$^{4}$}
\author{F.~Evers$^{1,2}$}

\affiliation{$^{1}$Institut f\"ur Nanotechnologie,
Forschungszentrum Karlsruhe, 76021 Karlsruhe, Germany \\
$^{2}$Institut f\"ur Theorie der kondensierten Materie, Universit\"at
Karlsruhe, 76128 Karlsruhe, Germany\\
$^{3}$School of Mathematical Sciences,
University of Nottingham, Nottingham NG7 2RD,  UK\\ 
$^{4}$Fakult\"at f\"ur Physik, Universit\"at
Karlsruhe, 76128 Karlsruhe, Germany }

\date{March 10, 2006}

\begin{abstract}
Two exact relations between mutlifractal exponents are shown to hold
at the critical point of the Anderson localization transition. The first
relation implies a symmetry of the multifractal spectrum linking the
multifractal exponents with indices $q<1/2$ to those with $q>1/2$.
The second relation connects the wave function multifractality to that
of Wigner delay times in a system with a lead attached.
\end{abstract}

\pacs{73.20.Fz, 72.15.Rn, 73.43.-f, 05.45.Df\\[-0.5cm]}

%% 73.20.Fz 	Weak or Anderson localization
%% 72.15.Rn 	Localization effects (Anderson or weak localization)
%% 73.43.-f 	Quantum Hall effects
%% 05.45.Df 	Fractals
%%
%%  Not included:
%% 05.70.Jk 	Critical point phenomena
%% 71.30.+h 	Metal-insulator transitions and other electronic
%%              transitions  

\maketitle

Anderson localization transitions in disordered electronic systems
represent a remarkable class of quantum phase transitions. Here we
understand the term ``Anderson transition'' in a broad sense,
including both the lo\-ca\-li\-za\-tion-de\-lo\-ca\-li\-za\-tion 
transitions and the
quantum Hall transitions between two phases with localized states. A
hallmark of these transitions is the multifractality (MF) of electronic wave
functions, describing their strong fluctuations at criticality
\cite{multifrac-reviews}:
the wave functions are characterized by a whole set of
fractal dimensions $D_q$ different from the spatial
dimensionality $d$. While at present the wave
function MF is routinely observed only in computer
simulations, rapidly developing imaging techniques allow to hope for
its forthcoming experimental observation.

In this Letter we point out two exact relations 
satisfied by the multifractal dimensions. The first of
these relations connects exponents with $q$ larger and
smaller than 1/2. The second relation links the multifractal indices
for the wave functions of a closed system to those for the Wigner
delay times characterizing the wave scattering from the same system
via an attached lead.

We begin by considering a functional relation for the distribution
function of the local density of states (LDOS) $\rho$,
\begin{equation}
\label{e1}
{\cal P}_\rho(\tilde{\rho})  = \tilde{\rho}^{-3} {\cal
P}_\rho(\tilde{\rho}^{-1}).
\end{equation}
Here $\tilde{\rho}$ is the LDOS normalized to its average value,
$\tilde{\rho} = \rho / \langle\rho\rangle$ (the normalization factor
is not critical and plays no role for our discussion). What is of
central importance here is the status of Eq.~(\ref{e1}).
Specifically, this formula is exact on the level of the non-linear
$\sigma$-model (NL$\sigma$M). It was derived for the first time in
Ref.~\cite{mf94} for the case of systems with broken time-reversal
invariance (unitary symmetry class). The reasons for its general
validity were revealed in Ref.~\cite{fyodorov04}, and an explicit
derivation for all standard Wigner-Dyson symmetry classes was
provided in Ref.~\cite{SFS05}.

Clearly, a mapping of a particular microscopic model of a disordered
system (e.g., continuous model of an electron moving in a random potential
with certain correlation function, or a discrete tight-binding
model, etc.) onto the NL$\sigma$M  is not exact. More
specifically, it is approximately valid in the case of weak disorder
and breaks down for strong disorder. Therefore, the relation
(\ref{e1}) has the same status. Nevertheless, we argue that a
relation between the anomalous multifractal exponents $\Delta_q$
characterizing the behavior of the moments $\langle\rho^q\rangle$ at
criticality,
\begin{equation}
\label{e2}
\Delta_q = \Delta_{1-q},
\end{equation}
which follows from (\ref{e1}), is {\it exact}.

Before explaining this, we digress with a brief reminder of the
wavefunction MF formalism; the reader is referred to
the reviews \cite{multifrac-reviews} for a more detailed exposition.
The moments of a wave function (so-called inverse participation
ratios) $P_q=\int d^d r |\psi({\bf r})|^{2q}$ show at criticality an
anomalous scaling with respect to the system size $L$,
\begin{eqnarray}
\label{e3}
&& \langle P_q \rangle = L^d \langle |\psi({\bf r})|^{2q}
\rangle \propto L^{-\tau_q}\ , \\
&& \tau_q \equiv D_q(q-1) = d(q-1) + \Delta_q.  \label{e3a}
\end{eqnarray}
Here $\Delta_q$ are anomalous exponents distinguishing a critical
point from a metallic phase [where $\tau_q=d(q-1)$].
%Here $\Delta_q$ are anomalous exponents distinguishing a critical
%point from a metallic phase [where $\tau_q=d(q-1)$] and related to
%the multifractal dimensions $D_q$ as $\Delta_q=(D_q-d)(q-1)$.
These exponents
also govern the scaling of the moments of LDOS,
\begin{equation}
\label{e4}
\langle \rho ^q \rangle \propto L^{-\Delta_q}.
\end{equation}
Equivalently, the
MF can be described by so-called singularity spectrum
$f(\alpha)$, which is the Legendre transform of $\tau_q$. Its
meaning is as follows: the average measure of a set of those points
$\bf r$ in a sample, where the wave function behaves as
$|\psi^2({\bf r})|\sim L^{-\alpha}$, scales with $L$ as
$L^{f(\alpha)}$.

We return now to the proof of exactness of
Eq.~(\ref{e2}). The central argument relies crucially on the
universality of critical properties at
the Anderson transition. Specifically, while the
mapping of the original microscopic model onto the NL$\sigma$M is
at most approximate, one can find another microscopic model (e.g.
$N$-orbital Wegner model in the limit $N\to \infty$ \cite{wegner},
which can also be viewed as a model of a granular metal
\cite{efetov-book}) that can be reduced exactly to the  
NL$\sigma$M. In view of the universality, the original
microscopic model and the NL$\sigma$M must flow into the same
fixed point in the infrared limit and will thus have the same
critical exponents. Therefore, the relation (\ref{e2}) must hold not
only in the NL$\sigma$M approximation, but be an intrinsic
property of any generic microscopic model, even though the validity
of Eq.~(\ref{e1}) is in general only approximate.

The moments of the LDOS and of the wave function intensity, which we
considered  above, are properties of a closed system. An
alternative method to study the local properties is to open the
system by attaching a perfectly coupled single-channel lead at a
point $\bf r$. The system can then be characterized by the Wigner
delay time $t_W$ (energy derivative of the scattering phase shift),
whose statistical properties attracted a lot of research interest in
recent years, see \cite{FS1997,ossipov05}. For convenience, we will
consider below the dimensionless delay time $\tilde{t}_W=t_W
\Delta/2\pi$ normalized to the mean level spacing $\Delta$. At the
Anderson transition point the corresponding distribution function,
${\cal P}_W(\tilde{t}_W)$, will reflect the criticality of the
system \cite{ossipov05,kottos05}.

To establish a connection between the MF of wave
functions and that of delay times, we recall a relation between
${\cal P}_W$ and the distribution function ${\cal P}_y$ of
normalized wave function intensities $y=V|\psi^2({\bf r})|$ 
($V\sim L^d$ is the system volume),
\begin{equation} \label{e5}
{\cal P}_W(\tilde{t}_W)=\tilde{t}_W^{-3} {\cal P}_y(\tilde{t}_W^{-1}).
\end{equation}
This formula was derived in Ref.~\cite{ossipov05} and has the same
status as Eq.~(\ref{e1}): it is exact on the level of the
NL$\sigma$M.  In particular, it implies the corresponding
relation between the exponents \cite{ossipov05}
\begin{equation}\label{e6}
    \gamma_q = \tau_{1+q},
\end{equation}
where the indices $\gamma_q$ characterize the scaling of moments of
the inverse delay time, $\langle t_W^{-q}\rangle \propto
L^{-\gamma_q}$. Applying the same argumentation as used above for
derivation of Eq.~(\ref{e2}), we conclude that the
relation (\ref{e6}) must again be {\it exact} for any generic
microscopic model.

The following point should be emphasized here.
Strictly speaking, the moments $\langle t_W^{-q}\rangle$ with
$q<-3/2$ are divergent for the one-channel scattering problem. To
define the exponent $\gamma_q$ for this case one should consider a
lead with several conducting channels. This is analogous to the
coarse-graining procedure for defining the wave-function 
exponent $\tau_q$ with negative $q$ discussed below. 
Equation (\ref{e6}) holds for such negative $q<-3/2$
as well, by analytical continuation. 

We turn  now to the analysis of consequences and applications of the
derived relations, mainly concentrating on Eq.~(\ref{e2}). First,
we rewrite the relation (\ref{e2}) in terms of the exponents
$\tau_q$,
\begin{equation}\label{e7}
    \tau_q-\tau_{1-q}=d(2q-1).
\end{equation}
Second, performing the Legendre transformation,
$f(\alpha_q)=q\alpha_q - \tau_q$ with $\alpha_q=d\tau_q/dq$, we get
\begin{equation}\label{e8}
    f(2d-\alpha) = f(\alpha) + d-\alpha.
\end{equation}
Equation (\ref{e8}) maps the part of the singularity spectrum with
$\alpha<d$ to that with $\alpha>d$. A particular consequence of this
is that the support of the singularity spectrum $f(\alpha)$ (i.e. the
region where
it is different from $-\infty$) is bounded by the interval $[0,2d]$.
The lower boundary, $\alpha\ge 0$, is a trivial consequence of the
wave function normalization; the upper boundary, $\alpha\le 2d$,
follows then from our relation (\ref{e8}).

It is worth mentioning that the results for the $f(\alpha)$
spectrum, as obtained numerically for the 3d Anderson transition in
a number of publications \cite{3d-multifractality,schreiber99}, are
in contradiction with this upper boundary [and thus with the
relation (\ref{e8})]. We believe that this is a consequence of an 
incomplete analysis of numerical data
in \cite{3d-multifractality,schreiber99}. 
Indeed, it was shown recently \cite{evers01} that the
earlier numerics on the wave function MF suffered
strongly from the absence of ensemble averaging and from finite-size
effects. The problems become even more severe for negative moments,
$q<0$, corresponding to the large-$\alpha$ part of the singularity
spectrum. This is evident, in particular, from Fig.~6 of
Ref.~\cite{schreiber99} where a strong drift of large-$\alpha$ part
of $f(\alpha)$ (towards our upper boundary $\alpha\le 2d=6$) with
increasing system size is seen.

Let us analyze the implication of our relation for the weak-coupling
expansion of the critical exponents that can be developed
in $2+\epsilon$ dimensions (where MF is weak). Since
Eq.~(\ref{e2}) is exact, it should hold in all orders of the
$\epsilon$-expansion. The known results for the
$\epsilon$-expansion of $\Delta_q$ up to 4-loop order \cite{wegner87}
do satisfy this property. In particular, the result for the orthogonal
symmetry class reads
\begin{equation}
\label{e9}
\Delta_q = q(1-q)\epsilon + {\zeta(3)\over
  4}q(q-1)[q(q-1)+1]\epsilon^4 + O(\epsilon^5).
\end{equation}
It is indeed seen that $\Delta_q$ depends on $q$ via the combination
$q(1-q)$ only, in agreement with the relation (\ref{e2}).

As a further application of Eq.~(\ref{e2}), we consider the model of
power-law random banded matrices (PRBM), $\langle|H_{ij}|^2\rangle =
(1+|i-j|^2/b^2)^{-1}$. This model (that describes a 1d system with
long-range $1/r$ random hopping) defines a family of critical
theories parametrized by $0<b<\infty$ and allows to study the
evolution of the critical system from the weak- to the
strong-MF regime with decreasing $b$
\cite{mirlin96,mirlin00}. While for $b\gg 1$ (weak MF)
the PRBM model can be approximately mapped to the NL$\sigma$M, 
for small $b$ (strong MF) this mapping
is not applicable, and the multifractal spectrum was analyzed in
\cite{mirlin00} by a different method. Our statement about the exactness
of Eqs.~(\ref{e2}), (\ref{e6}) remains valid for the PRBM model.
Indeed, we can construct a ``granular'' generalization of the model
with $N\gg 1$ states at each site of the 1d lattice and with hopping
matrix elements between all states decaying with distance $r$ as
$(\tilde{b}/N)r^{-1}$. Changing the overall prefactor $\tilde{b}$ in
the hopping amplitude will yield a family of critical models that
should flow in the infrared limit to the same line of critical
points as the family of PRBM models. In this way,  the PRBM model
with an arbitrary value of $b$ can be associated with an N-orbital model
with some $\tilde{b}$ that will have the same critical properties.
On the other hand, the latter model can be mapped onto the
NL$\sigma$M, which allows us again to derive the relations
(\ref{e2}), (\ref{e6}) for the critical exponents.

We have verified the validity of the relation (\ref{e2}) by a
numerical simulation of the PRBM model. The exponents $\tau_q$ were
extracted from the scaling of the inverse participation ratios $\langle
P_q\rangle$ for system sizes $L$ in the range  from 512 to 4096.
The number of disorder realizations was ranging from $2\times 10^5$
for $L=512$ to $1000$ for $L=4096$. It should be stressed that
evaluation of negative moments requires special care, since the
inverse participation ratio, as defined in Eq.~(\ref{e3}), is
divergent because of zeros of the wave function. These zeros, related
to oscillations of the wave function on the scale of the wave length,
have nothing to do with multifractal properties characterizing smooth
envelopes of wave functions. To find $\tau_q$ with negative $q$, we 
have first smoothed $|\psi^2|$ by averaging over blocks of the size
$m=16$, and then applied Eq.~(\ref{e3}). This makes finite-size effects
(and thus a numerical inaccuracy in evaluation of $\tau_q$)
for negative moments considerably more pronounced than for $q>0$. 

\begin{figure}
\begin{center}
\includegraphics[width=0.9\columnwidth]{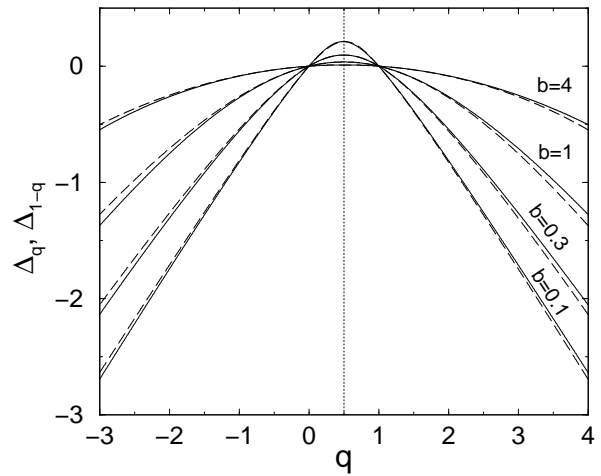}
\vspace*{-0.5cm}
\caption{Multifractal exponents $\Delta_q$ for the PRBM model with
$b=4$, 1, 0.3, 0.1. The symmetry (\ref{e2}) with respect to the point
$q=1/2$ 
is evident. A small difference between $\Delta_q$ (full
line) and $\Delta_{1-q}$ (dashed) is due to numerical errors.}
\label{fig1}
\end{center}
\vspace*{-0.5cm}
\end{figure}

\begin{figure}
\begin{center}
\includegraphics[width=0.9\columnwidth]{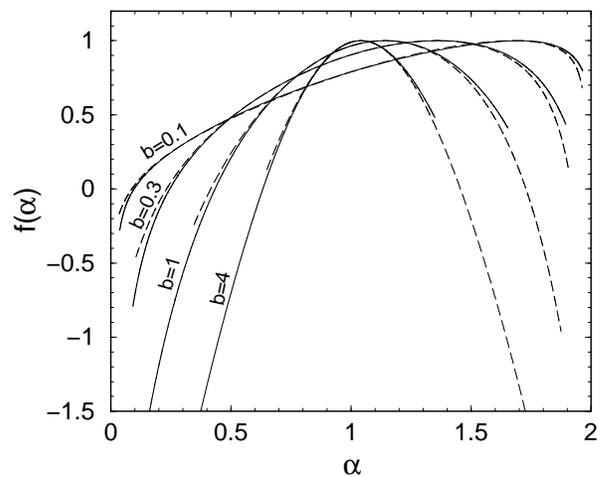}
\vspace*{-0.5cm}
\caption{The data of Fig.~\ref{fig1} in terms of the singularity
  spectrum $f(\alpha)$. Dashed lines represent $f(2-\alpha)+\alpha-1$,
  demonstrating the validity of Eq.~(\ref{e8}).}
\label{fig2}
\end{center}
\vspace*{-0.5cm}
\end{figure}

The results of the numerical simulations for the PRBM ensemble with
several values of $b$, spanning the whole interval from the
weak-MF to strong-MF regime,
are shown in Figs.~\ref{fig1}, \ref{fig2}. The data in Fig.~\ref{fig1}
nicely confirm the symmetry relation (\ref{e2}). A small difference
between $\Delta_q$ and $\Delta_{1-q}$ can be considered as a measure
of the numerical accuracy of evaluation of the exponents. As discussed
above, the errors are mainly due to moments with negative $q$. 
In Fig.~\ref{fig2} the same numerical data are presented in the form of the
singularity spectrum $f(\alpha)$. To demonstrate that the data support
very well the relation (\ref{e8}), we also show the function
$f(2-\alpha)+\alpha-1$.  

\begin{figure}
\begin{center}
\includegraphics[width=0.9\columnwidth]{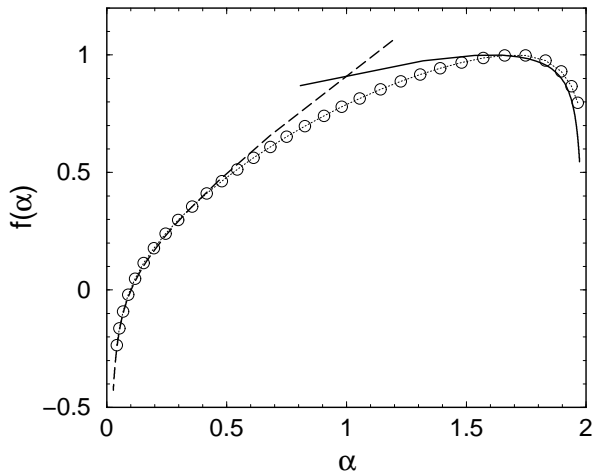}
\vspace*{-0.5cm}
\caption{Singularity spectrum for the PRBM model with $b=0.1$.
Dashed line: $\alpha<1$ behavior, Eq.(\ref{e12}); full line:
$\alpha>1$ result, Eq.~(\ref{e14}), following from the relation
(\ref{e8}); circles: numerical data. Some
mismatch between the slopes of the two curves at $\alpha=1$ is
related to the fact that the formula (\ref{e12}) is valid to the
leading order in $b\ll 1$.}
\label{fig3}
\end{center}
\vspace*{-0.5cm}
\end{figure}

We will now demonstrate the high utility of Eq.(\ref{e2}) by
applying it for the analytical evaluation of exponents with $q<1/2$ 
in the "non-NL$\sigma$M" limit, $b\ll 1$. As was
found in \cite{mirlin00}, the multifractal exponents in this regime
are given for $q>1/2$ by
\begin{eqnarray}
\label{e10}
  \tau_q & \simeq & 2b T(q)\ ,\\[0.3cm]
  T(q) &=& {2\over \sqrt{\pi}} {\Gamma(q-1/2)\over \Gamma(q-1)}
  \nonumber \\
  & \simeq & \left\{ \begin{array}{ll}
  -{1\over \pi (q-1/2)}\ , \qquad & q \to 1/2, \\
{2\over \sqrt{\pi}} q^{1/2}\ , \qquad & q\gg 1.
\end{array}
\right. \label{e11}
\end{eqnarray}
In terms of the singularity spectrum $f(\alpha)$, this means
\begin{equation}\label{e12}
    f(\alpha)\simeq 2bF(\alpha/2b),
\end{equation}
where $F(A)$ is the Legendre transform of $T(q)$ with the
asymptotics
\begin{equation}\label{e13}
    F(A)\simeq \left\{ \begin{array}{ll}
    -1/\pi A\ , \qquad & A\to 0\ , \\
    A/2\ , \qquad & A\to\infty\ .
    \end{array}
    \right.
\end{equation}
Equation (\ref{e10}), (\ref{e11}) was derived in \cite{mirlin00} by
a real-space renormalization-group method valid for $q>1/2$. The
relation (\ref{e2}) allows us now to find the multifractality
spectrum for $q<1/2$. When translated to $f(\alpha)$-language, this
yields the singularity spectrum for $\alpha>1$,
\begin{eqnarray}
 f(\alpha) &=& f(2-\alpha)+\alpha-1 \nonumber \\
  & \simeq & 2b F\left({2-\alpha\over 2b}\right) +\alpha -1\nonumber\\
  &\simeq & \left\{ \begin{array}{ll}
    \alpha/2\ , \qquad & 2-\alpha\gg 2b\ , \\
    1-{4b^2\over \pi(2-\alpha)}\ , \qquad & 2-\alpha\ll 2b\ .
    \end{array}
    \right. \label{e14}
\end{eqnarray}
In Fig.~\ref{fig3} we show the MF spectrum of the PRBM
model for $b=0.1$. The dashed curve yields the $\alpha<1$ behavior,
Eq.(\ref{e12}), while the full line is the $\alpha>1$ result,
Eq.~(\ref{e14}). 

%%%%%%%%%%%%%%%%%%%%%%%%%%%%%%%%%%%%%%%%%%%%%%%%%%%%%%%%%%%%%%%%
%%%%%%%%%%%%%%%%%%%%%%%%%%%%%%%%%%%%%%%%%%%%%%%%%%%%%%%%%%%%%%%%

%%%%%%%%%%%%%%%%%%%%%%%%%%%%%%%%%%%%%%%%%%%%%%%%%%%%%%%%%%%%%%%%
%%%%%%%%%%%%%%%%%%%%%%%%%%%%%%%%%%%%%%%%%%%%%%%%%%%%%%%%%%%%%%%%

%%%%%%%%%%%%%%%%%%%%%%%%%%%%%%%%%%%%%%%%%%%%%%%%%%%%%%%%%%%%%%%%
%%%%%%%%%%%%%%%%%%%%%%%%%%%%%%%%%%%%%%%%%%%%%%%%%%%%%%%%%%%%%%%%

In the limit $b\to 0$ the MF reaches its extreme form
(for the PRBM model the effective spatial dimensionality $d=1$; we
keep $d$ below for generality)
\begin{equation}\label{e15}
    \tau_q=\left\{ \begin{array}{ll}
    0\ , \qquad & q\ge 1/2\ , \\
    d(2q-1)\ , \qquad & q\le 1/2\ ,
    \end{array}
    \right.
\end{equation}
or, in terms of the singularity spectrum,
\begin{equation} \label{e16}
f(\alpha)= \left\{ \begin{array}{ll} \alpha/2\ , \qquad & 0\le
\alpha \le 2d\ , \\
-\infty \ , \qquad & {\rm otherwise} \ .
\end{array}
    \right.
\end{equation}

The following remark is in order here. The earlier analysis
\cite{mf94,mildenberger02} of the statistics of critical wave
functions on the Bethe lattice and in large dimensionality
$d$ allows us to conjecture that in the limit $d\to\infty$ the
multifractal spectrum at the Anderson transition acquires the same
extreme form (\ref{e15}), (\ref{e16}).  We stress,
however, that this is only a hypothesis waiting for a more rigorous
verification.

The second relation we claim to be exact, Eq.(\ref{e6}), is also
supported by numerical results obtained for the PRBM model.
Specifically, the numerical data \cite{kottos05} for the scaling of
the delay time moments confirm (in combination with the results of
Ref.~\cite{mirlin00} on the wave function MF) the
validity of Eq.~(\ref{e6}) even in the small-$b$ limit where the
mapping of the PRBM to the NL$\sigma$M fails.

As a final remark, we note that the notion of MF was
recently extended to the surface of a critical system
\cite{surface-multifrac}. While boundary multifractal exponents are
different from their bulk counterparts, the relations (\ref{e2}) and
(\ref{e6}) remain valid also for surface MF. Indeed, it
is not difficult to check that the derivation of the relations for
the distribution functions (that served as starting points for our
analysis), (\ref{e1}) and (\ref{e5}), retain its validity
independently on the position (in the bulk or near the boundary) of
the observation point ${\bf r}$. The MF of delay times
for a lead attached to the boundary has in fact been studied
numerically in the PRBM model in Ref.~\cite{kottos05}; 
an analysis of the surface MF of wave functions and the
verification of the relation (\ref{e6}) 
at the boundary of this system will be
presented elsewhere. 

To summarize, we have demonstrated that two exact relations,
Eq.(\ref{e2}) [that can be equivalently represented in the form
(\ref{e7}) or (\ref{e8})] and Eq.(\ref{e6}), hold for multifractal
exponents at the critical point of the Anderson transition. We have
applied the first of these relations to the multifractality spectrum
of the PRBM model and verified its validity by numerical simulations.
A further analysis of implications of these
relations is of
considerable interest. Another direction of future research is to
study whether these relations, derived here for three Wigner-Dyson
classes, have some analogues for unconventional (chiral and
superconducting) symmetry classes \cite{altland-zirn}.

We thank I.~Gruzberg
for useful discussions. This work was supported by the
SPP ``Quanten-Hall-Systeme'' and the Center for Functional
Nanostructures of the DFG (ADM, FE), and by EPSRC grant EP/C515056/1 (YF).

\end{document}